# A Big Data Analysis of the Ethereum Network: from Blockchain to Google Trends

Dorsa Mohammadi Arezooji

*Center for Complex Networks and Social DataScience (CCNSD)*
*Department of Physics, Shahid Beheshti University*
*Tehran, Iran*

d.mohammadiarezooji@se19.qmul.ac.uk

*Abstract*: First, a big data analysis of the transactions and smart contracts made on the Ethereum blockchain is performed, revealing interesting trends in motion. Next, these trends are compared with the public's interest in Ether and Bitcoin, measured by the volume of online searches. An analysis of the crypto prices and search trends suggests the existence of big players (and not the regular users), manipulating the market after a drop in prices. Lastly, a cross-correlation study of crypto prices and search trends reveals the pairs providing more accurate and timely predictions of Ether prices.

## 1. Introduction

Blockchains technology and cryptocurrencies such as Ethereum and Bitcoin have attracted the rapidly increasing attention of many industries and researchers from various fields. Part of the Blockchain's appeal to physicists and data scientists stems from its complexity and underlying interdependencies that affect and take influence from other complex systems namely financial and social networks.

The intuition behind bringing social networks into consideration while analyzing blockchains, lies within their decentralized structure. At its core, decentralization was introduced to remove heterogeneity from systems. In the case of cryptocurrencies, this would cause governments and central banks to lose their monopoly in financial markets. This would also mean that the network itself (active users) now controls the changes in the blockchain. Needless to say, the power of decentralization depends on how decentralized the network really is. In other words, in the case of disproportionately weighted "central" nodes, the network would be prone to severe heterogeneity threats [1]. Several studies have explored the degree of centrality in decentralized financial networks, revealing that they may not be as decentralized as one would expect [2].

## 2. Background

The initial idea that led to the birth of blockchain was first proposed in 1990 by S. Haber and W. S. Stornetta. They introduced a novel, computationally feasible set of procedures to timestamp digital data, which would make it impractical to alter the timestamp after its creation [3]. One of the most important implications of this approach is that there would be no need for a third-party service to keep record of the timestamps. Since then, blockchain technology has revolutionized many fields including healthcare [4,5], transportation [6,7], digital forensics [8], and cybersecurity [9,10] due to its reliability, immutability, and transparency. These characteristics are the direct results of the blockchain structure: data is divided into a collection of blocks that are all linked together by means of cryptography. This structure prevents tampering with any arbitrary block without changing all others; hence, achieving immutability. Furthermore, the data stored in any node across the network is visible to all users, thus, transparency is maintained. The decentralized data handling also prevents the two parties in a transaction to retroactively manipulate data stored in the network. In a nutshell, the blockchain establishes a general agreement that verifies the details of a transaction without the need to trust the parties involved [11].

### 2.1. The Ethereum Blockchain

The Ethereum blockchain is a type of distributed ledger technology (DLT), a general term used for databases that store and share information in a decentralized network of independent nodes. Even though most terms used in the context of Ethereum are not exclusive to this particular blockchain, there are a few terms that are. This section provides a brief introduction to the technical terms related to the Ethereum blockchain that are used in this paper. A comprehensive explanation of terms and concepts can be found in the Ethereum whitepaper (ethereum.org).

#### 2.1.1. Gas

On the Ethereum blockchain, the cost of performing transactions or processing smart contracts is measured by gas. The price of gas itself is not constant, but is reported by miners based on the complexity and computational resources required for the execution of each block. Gas fees are calculated in ether (ETH), which is Ethereum's native currency. The smallest denomination of ether is named wei (1e-18 ETH). Gas price is usually reported in Gwei (1e-9 ETH).

#### 2.1.2. Smart Contract

The concept of smart contracts was introduced by N. Szabo in 1994 in an unpublished manuscript and then formally in 1997 [12]. Smart contracts are essentially blockchain-based applications, concisely, self-executing programs which contain the terms and agreements between the parties involved. The feature that sets smart contracts apart is that they automatically verify whether the terms of the agreement have been fulfilled or not. Additionally, to ensure reliability, fault tolerance, and transparency, the codes (smart contracts) are replicated on many nodes in the blockchain. Interestingly, the Ethereum blockchain has hosted roughly 1.5 million smart contracts in the last few years [13].

*2.2. Big Data Processing*

Big data is a term used to describe data that is too large to be processed with conventional methods, and keeps growing exponentially with time. Big data processing tools and frameworks such as Apache Hadoop enable distributed processing of large-scale data, a task made possible by a network of heterogeneous computation units. In essence, a Hadoop cluster can be a cluster of commodity PCs (or virtual machines) that are connected and communicate with a master node. The philosophy of Hadoop is "moving the computation to the data". This entails each node serving as both a storage unit and a processing unit. These two components, HDFS (Hadoop Distributed File System - storage) and YARN (Yet Another Resource Negotiator - computation), have been designed to work together in a single cluster [14,15]. By distributing the data and performing the computation in parallel, Hadoop is able to achieve scalability from a single node to thousands of nodes. Hadoop uses MapReduce to distribute the data among worker nodes. Similarly to the blockchain, the data is split into a number of blocks, each assigned to a node across the cluster.

Apache Spark is another big data framework which is built on top of HDFS. By reducing disk read and write operations, Spark provides fast in-memory processing [16]. It is also suitable for both batch and streaming data, making it a versatile tool for a vast range of big data processing tasks. These tasks include machine learning, graph processing, and interactive queries on big data. Finally, Spark is written in Scala but has APIs in Java, R, and Python.

*2.3. The Social Factor*

As it was previously mentioned, the interaction between complex networks such as the Ethereum blockchain and social networks can be quite fascinating. Analyzing social factors may offer an opportunity to gain insight into the events that induce changes in the market. With this in mind, Google Trends is chosen as a data source for assessing public interest in cryptocurrencies such as Ether and Bitcoin. The data shows the public interest measured by the aggregated sum of online searches. Google Trends data is publicly available with a few restrictions on data frequency.

## 3. Data and Tools

The Ethereum dataset used for this analysis was originally collected from the dumps uploaded to a repository on Google's BigQuery, now available as a public dataset. The dataset was uploaded to a Hadoop cluster and stored in HDFS.

Due to its high speed and performance, Spark (in this study, specifically its python API, PySpark) was used to process the dataset (in order of TBs). The monthly total number of transactions, average gas, and gas price were aggregated and extracted from the dataset.

## 4. Analyses and Results

An initial analysis of the data is visualized in fig 1, revealing a surge in the number of transactions, including smart contracts, in early 2018. Interestingly, this event coincides with the sudden surge in Ether and Bitcoin prices. The possible links between Ether and Bitcoin prices will be discussed in more depth. Another observation points to a rather steady decrease in gas price, with a slight increase in the beginning or at the end of each year. In addition, the average gas used for transactions seem to have reached a steady state as of late 2017. Finally, since with more complex contracts, more gas would be needed, the strong correlation (pearson correlation coefficient of 0.94) between smart contract's difficulty and required gas can be explained.

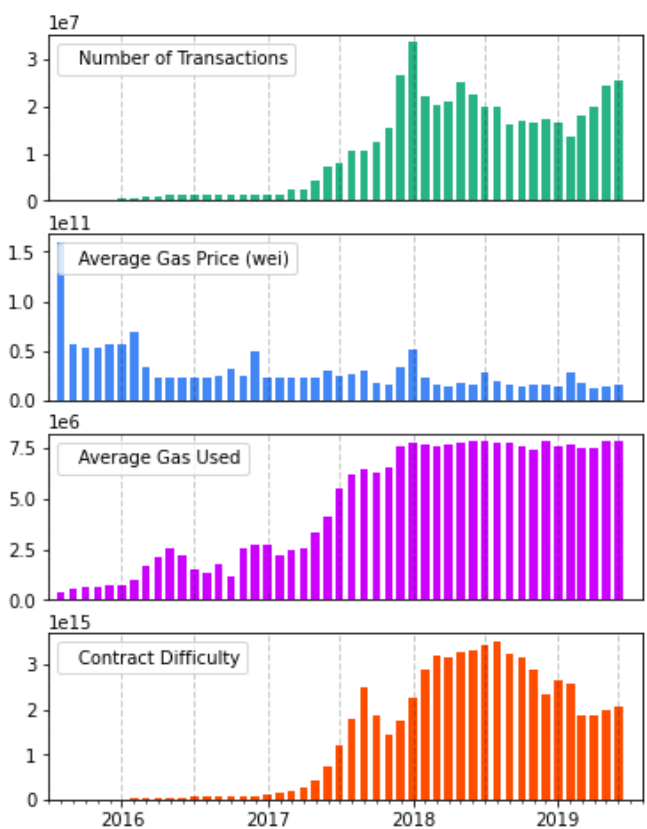

Fig. 1. Ethereum contracts and transactions over time

Moving forward, Google Trends' data is incorporated into the analysis as an exogenous feature. The first logical notion could be that with more people searching Ethereum online, more of them might invest and make transactions on the network. Hence, the number of transactions and the public interest in Ethereum are depicted in fig 2. It should be noted that the y axis is indicative of only the number of transactions and not the volume of online searches. Daily Google Trends data have been collected in 180-day intervals, then scaled and concatenated and finally normalized on the scale of 100. For a better visualization, they have been scaled once more to fit the y axis of the plot. The final results show that following the first abrupt increase in Google Trends' data in mid 2017, the number of transactions didn't experience the same dramatic change, possibly pointing to a lack of trust in the cryptocurrency and the blockchain. In any case, in early 2018 the number of transactions on the platform reached its highest as the volume of online searches also hit an all-time high. A rudimentary conclusion could be that an increase in the public's interest in a cryptocurrency corresponds with an increase in the number of transactions made. Nevertheless, perhaps a more enticing objective is utilizing this information to predict how these factors affect the price of ETH.

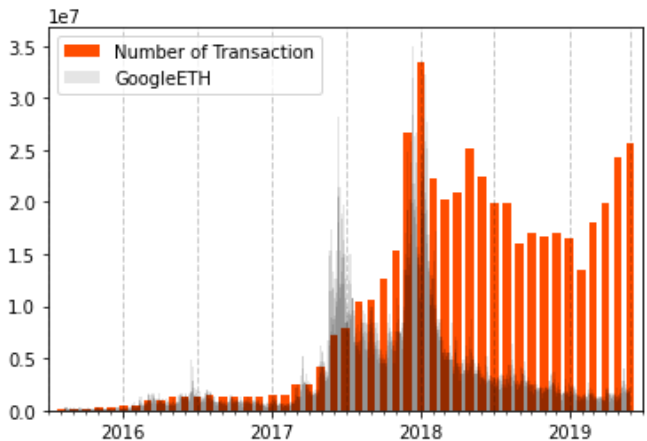

Fig. 2. Number of transactions vs. online search trends

In line with the same analogy as above, the price of ETH and Bitcoin (BTC) are plotted in fig 3, along with their respective search volumes. It is speculated that the surge in ETH/USD in 2018 was correlated with the rise in BTC/USD, which itself was triggered by Tether (USDT). A study of the Bitcoin blockchain by Griffin and Shams showed that Tether-based transactions on the Bitcoin blockchain were timed following the drop in BTC/USD, setting in motion the soaring of Bitcoin prices [17].

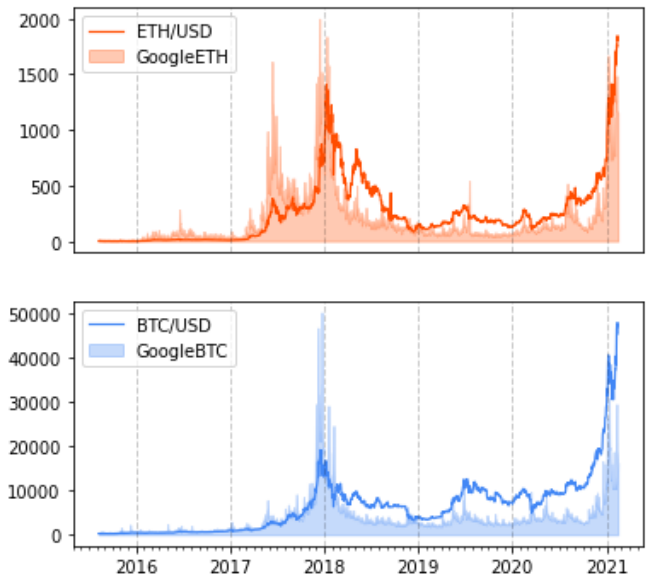

Fig. 3. Cryptocurrency prices vs. online search trends

At first glance, a side by side analysis of the crypto pair shows a similar trend. Both ETH/USD and BTC/USD skyrocketed in early 2018, following a rise in the public's interest in the crypto market. However, after the similarly sharp fall in search volumes during 2018, crypto prices (especially ETH) did not drop as expected according to the public's interest. This could be indicative of the existence of another entity other than the ordinary users entering the market: whales (large and powerful entities) holding on until after the drop to invest heavily. This assumption is in accordance with the conclusions made by Griffin and Shams, stating that crypto prices are not solely influenced by supply and demand [17]. Their analysis suggests that the distortion in Bitcoin prices was caused by a large player on Bitfinex using Tether to invest in large sums of Bitcoin.

Considering the mechanisms responsible for the change in crypto prices, a statistical analysis of their returns can prove to be beneficial in deciding which assets to invest in. To obtain a stationary time series representing the changes in crypto prices, logarithmic returns of ETH/USD and BTC/USD are computed and plotted against time in fig 4.

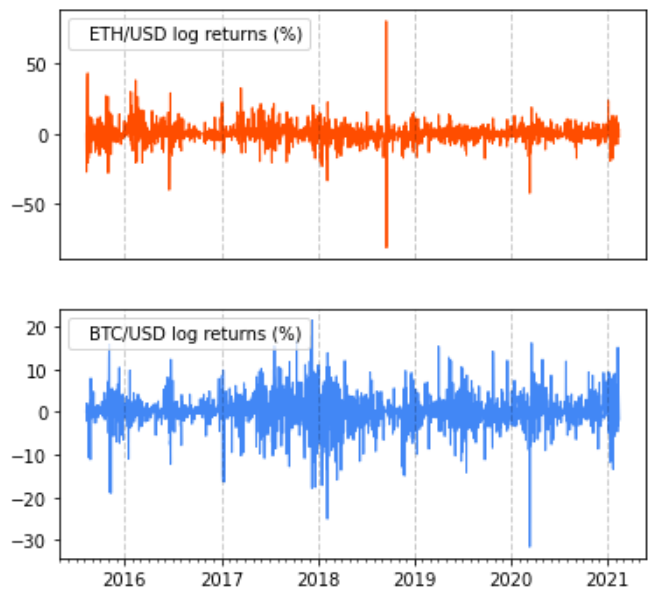

Fig. 4. Cryptocurrency log returns over time

Comparing the probability density function (PDF) of Ether and Bitcoin's log returns (fig 5), it can be deduced that both seem to follow a gaussian distribution. However, Bitcoin's log returns' PDF appears to be narrower and more peaked than Ether, hence, it would be a more suitable choice for risk-averse investments.

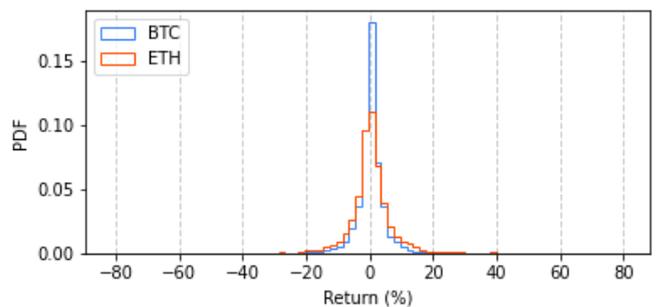

Fig. 5. Cryptocurrency log returns histogram

A more descriptive and quantifiable summary of the log returns are available in table 1. Interestingly, Ether's log returns hit a maximum of 80%, almost 3 times higher than that of Bitcoin. Nonetheless, Ether's minimum log return had dropped to -81%, about 1.5 times lower than that of Bitcoin.

Table 1. Statistical summary of prices and log returns

|  | *ETH/USD* | *BTC/USD* | *%R<sub>ETH</sub>* | *%R<sub>BTC</sub>* |
|---|---|---|---|---|
| **Mean** | 247.30 | 6355.57 | 3.38 | 0.25 |
| **std** | 284.43 | 6588.89 | 6.90 | 4.02 |
| **Min.** | 0.43 | 203.18 | -81.26 | -31.59 |
| **1st Qu.** | 12.83 | 821.19 | -2.27 | -1.15 |
| **2nd Qu** | 182.87 | 6148.42 | 0.12 | 0.23 |
| **3rd Qu** | 323.72 | 9177.89 | 2.87 | 1.86 |
| **Max.** | 1845.82 | 47884.18 | 80.40 | 21.45 |

Finally, to understand how long it takes for changes to be reflected in crypto prices, the cross-correlation between the 4 time series are calculated. The time lag which corresponds to the maximum cross-correlation, denoted as "lag max", shows the number of days until the second time series is maximally correlated with the first. Figure 6 illustrates a heatmap of the maximum cross-correlation between each pair of the four time series. The highest cross-correlation scores belong to the (ETH/USD, BTC/USD), (GoogleETH, GoogleBTC), and (ETH/USD, GoogleBTC) pairs respectively, suggesting that these pairs could serve well as predictors.

Fig 6. Max cross-correlations heatmap

|  | ETH/USD | GoogleETH | BTC/USD | GoogleBTC |
|---|---|---|---|---|
| ETH/USD |  | 0.75 | 0.84 | 0.79 |
| GoogleETH | 0.75 |  | 0.59 | 0.83 |
| BTC/USD | 0.84 | 0.59 |  | 0.67 |
| GoogleBTC | 0.79 | 0.83 | 0.67 |  |

The changes in cross-correlations are visualized against time lag in fig 6. It can be observed that the longest time lag belongs to the (ETH/USD, GoogleBTC) pair. In simple terms, this indicates that an increase in the volume of online searches for Bitcoin correlates with an increase in the ETH/USD approximately a month later. Based on time lags, the (ETH/USD, GoogleBTC), (ETH/USD, BTC/USD), and (ETH/USD, GoogleETH) pairs are the strongest contenders for timely predictions.

It should be noted that in order to use this information for future analyses and prediction purposes, the strength of cross-correlations should be taken into consideration as well. Thus, the (ETH/USD, BTC/USD) and (ETH/USD, GoogleBTC) pairs provide both strong and timely predictions compared to other pairs. Finally, it is crucial to bear in mind that while the above statements are centered around cross-correlation, this concept should not be confused with causation. Studying the possible causal links requires utilizing other methods such as Bayesian structural learning and a deep understanding of economics and Blockchain.

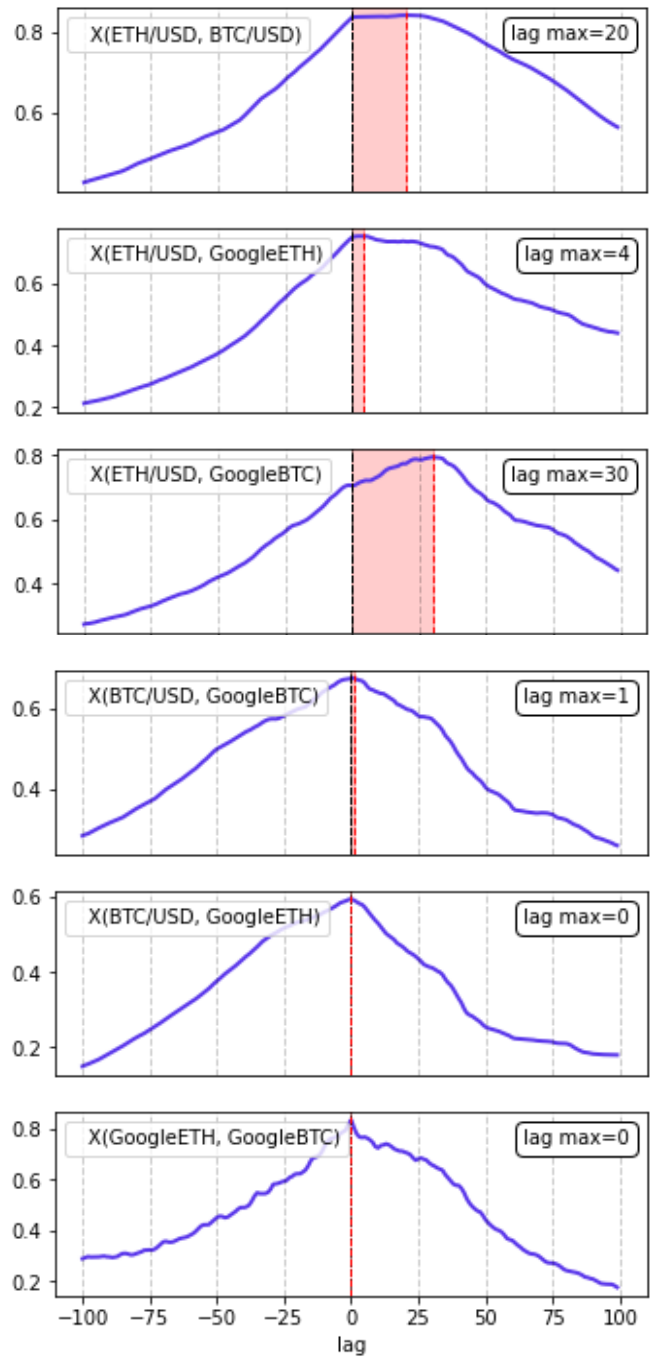

Fig. 7. Cross-correlations vs. time lag

## 5. Discussion and Conclusions

The analyses in this study suggest that considering the couplings between social and financial complex systems helps making better predictions about asset prices. Although here the links between only two assets were explored, it would be beneficial to extend this analysis to a wider range of components. Furthermore, this approach helps identify market manipulations and crypto price bubbles caused by entities other than the public. Future analyses may incorporate other sources of social data such as twitter to gain more insight.